\documentclass[journal=jacsat,manuscript=article]{achemso}
\usepackage{chemformula} 
\usepackage[T1]{fontenc}
\usepackage{graphicx}
\usepackage{amsmath}

\usepackage{amssymb}
\usepackage{bm}
\usepackage{caption}
\usepackage{subcaption}
\usepackage{hyperref}
\usepackage{nameref}
\usepackage{booktabs}
\usepackage{multirow}
\usepackage{wrapfig}
\usepackage{comment}
\DeclareMathOperator{\Tr}{Tr}
\usepackage[normalem]{ulem}
\usepackage[version=4]{mhchem}

\hypersetup{pdfborder=0 0 0,colorlinks=true,citecolor=black,linkcolor=black}
\input epsf

\newcommand{\editorr}[2]{%
  \expandafter\newcommand\csname #1note\endcsname[1]{%
    \textcolor{#2}{(\textbf{#1:} ##1)}}%
  \expandafter\newcommand\csname #1\endcsname[1]{%
    \textcolor{#2}{##1}}%
  \expandafter\newcommand\csname #1cancel\endcsname[1]{%
    \textcolor{#2}{\sout{##1}}}%
  \expandafter\newcommand\csname #1change\endcsname[2]{%
    \textcolor{#2}{\sout{##1} ##2}}%
  \newenvironment{#1text}{\color{#2}}{\color{black}}
}

\editorr{LB}{red}
\editorr{CC}{violet}

\newcommand\THEOSMARVEL{Theory and Simulation of Materials (THEOS), and National Centre for Computational Design and Discovery of Novel Materials (MARVEL), {\'E}cole Polytechnique F{\'e}d{\'e}rale de Lausanne, 1015 Lausanne, Switzerland}
\newcommand\BREMEN{U Bremen Excellence Chair, Bremen Center for Computational Materials 
Science, and MAPEX Center for Materials and Processes, Universit{\"a}t Bremen, 28359 Bremen, Germany}
\newcommand\SAPIENZA{Departement of Physics, Sapienza Universit{\`a} di Roma, Rome, Italy}
\newcommand\PSI{Laboratory for Materials Simulations, Paul Scherrer Institut (PSI), 5232 Villigen, Switzerland}

\author{Lorenzo Bastonero} \affiliation{\BREMEN} \email{lbastone@uni-bremen.de}
\altaffiliation{These authors contributed equally to this work}
\author{Chiara Cignarella} \affiliation{\BREMEN}
\altaffiliation{These authors contributed equally to this work}
\author{Nicola Marzari} \affiliation{\BREMEN} \alsoaffiliation{\THEOSMARVEL} \alsoaffiliation{\PSI}
\author{Lorenzo Monacelli} \affiliation{\SAPIENZA}

\title{Extreme anharmonicity determines the superionic transition in the prototypical antifluorite Li$_2$O}

 %5-7
 
\begin{document}

%------------------------------------------------------------------------------------ %
%------------------------------------------------------------------------------------ %
%TC:ignore
\begin{abstract}
\noindent Superionic conductors are promising electrolyte candidates for all-solid-state batteries, thanks to their liquid-like ionic mobility. In the prototypical lithium-ion oxide superionic conductor with antifluorite structure, the high-temperature transition to the fast-ion regime has long been attributed to volume-driven harmonic phonon softening. 
However, this quasi-harmonic picture neglects the strong lattice anharmonicity inherent to the fast-ion regime, leaving the microscopic origin of the transition unresolved.
Here, we apply non-perturbative lattice dynamics combined with a first-principles-accurate machine-learning interatomic potential to study the phase transition in \ce{Li2O}. We demonstrate that anharmonicity, rather than volume expansion alone as commonly assumed, plays the decisive role in determining the transition temperature and the underlying mechanism. Finally, we analyze the spectral properties using the time-dependent self-consistent harmonic approximation, revealing the breakdown of the quasi-particle picture in this material.
The spectral function and the infrared response provide signtures of the phase transition that can be investigated experimentally using inelastic neutron or X-ray scattering and infrared spectroscopy.
All these results establish the extreme anharmonicity of \ce{Li2O} near the superionic regime and provide a lattice-dynamic framework for understanding and engineering the fast-ion transition in this type of materials.
\end{abstract}
%TC:endignore
%------------------------------------------------------------------------------------ %
%------------------------------------------------------------------------------------ %
%--------------------- INTRODUCTION ------------------------------------------------- %
\vspace{5mm}%
\noindent \ce{Li2O} is a very interesting material for energy applications, already investigated in fusion technologies~\cite{Donato1998}, lithium-air batteries~\cite{Kwak2020} and solid-state electrolytes~\cite{Zhang2023}. 
It belongs to the family of antifluorite-type structures, including materials capable of conducting ions such as Li, F, and even Cu\cite{Wang2025b}. \ce{Li2O} has the \emph{Fm3m} space group and a face-centered cubic structure, with Li atoms occupying the tetrahedral sites.
The family of fluorites and antifluorites have received increasing attention thanks to their superionic phase.
Superionic conductivity is, in fact, a regime of fast-ion diffusion characterized by ionic conductivity in the range of that of liquids, where ions diffuse through a fixed sublattice, making superionic materials extremeling appealing for solid-state electrolytes~\cite{Wang2015}.
However, in fluorites and antifluorites, this phase usually occurs at too high temperatures to be used in real applications \cite{Farley1991, Goel2004, Gupta2012}, and finding new ways to engineer the sublattice chemistry would be highly beneficial~\cite{Chen2024b}. 
The superionic phase transition can occur as first-order (type-I) phase transition, characterized by a sudden change of the diffusivity above the critical temperature T$_c$, and as second-order (type-II), where the diffusion coefficient changes continuosly.
Li$_2$O belongs to the second group, and undergoes to a second-order phase transition with a continuous volume change, a specific heat anomaly and a large decrease in the elastic constant C$_{11}$ at the superionic transition temperature \cite{Goel2004,Gupta2012}.
Previous studies have suggested that Li-ion transport in \ce{Li2O} involves solid-like vibrations around lattice sites combined with hopping between sites, even in the superionic phase~\cite{Fracchia1998,Goel2004,Goel2007,Gupta2012}, similar to its sulfide counterpart \ce{Li2S}~\cite{Wang2015}.
The superionic phase transition in \ce{Li2O} is associated with a phonon softening occurring at a critical volume \cite{Gupta2012, Wang2015, Gupta2019}. In this view, the volume becomes the main driving force of the transition.
A similar mechanism was found in a high-pressure phase of \ce{CaF2}~\cite{Nelson2018} and verified using the quasi-harmonic approximations (QHA), where the calculated temperature associated with the softening was comparable with experiments. 
QHA calculations for \ce{Li2O}~\cite{Fracchia1998,Goel2004,Gupta2012}, however, overestimate the predicted critical volume with respect to the experimental volume of the transition, even when using different DFT functionals (LDA, PBE, PBEsol)~\cite{Nelson2018}, leading to assigning the choice of the latter as the main culprit of the mismatch. 
On the other hand, the core role of anharmonicity in other type of superionic materials was demonstrated both experimentally and computationally, by different independent works \cite{Niedziela2018,Ding2020,Gupta2021,Gupta2022,Ren2023,Ding2025,Buckeridge2013}, and neutron scattering experiments indicate an increasing anharmonicity towards the superionic transition temperature also in Li$_2$O \cite{Gupta2019}.
Studying lattice dynamics of superionic materials using the quasi-harmonic approximation~\cite{Fracchia1998,Goel2004,Gupta2012} does not correctly account for this strong anharmonicity in the superionic phase. In addition, the harmonic free energy is ill-defined in the presence of unstable modes~\cite{Baroni2001,Togo2015} ($\omega^2<0$) which accompany second-order structural phase transitions.
To date, an accurate lattice-dynamical description of the superionic transition in the prototype Li$_2$O that accounts for anharmonicity is still lacking.
Here, for the first time, we employ the stochastic self-consistent harmonic approximation (SSCHA) \cite{Monacelli2021a} to investigate Li$_2$O in the superionic regime.
We provide a comprehensive analysis of the lattice dynamics, fully accounting for anharmonic effects, to clarify the role of anharmonicity and elucidate the mechanism underlying the superionic phase transition.
SSCHA is a method to compute the quantum free energy of crystals and molecules that goes beyond the harmonic approximation and fully accounts for anharmonic and quantum effects in a non-perturbative fashion\cite{Monacelli2021a}. The free energy is computed by minimizing the functional F[$\rho$] of a trial density matrix $\rho$ of Gaussian-shape with respect to two parameters: the average atomic position $\mathcal{R}_0$ and the effective interatomic force constant (IFC) $\Phi$, corresponding to the centroids and the covariance matrix of the Gaussian.
The SSCHA method is particularly suited to study second-order phase transitions, as the free energy is well defined even when a phase is dynamically unstable~\cite{Monacelli2021a}.
We take advantage of a previously developed machine-learning interatomic potential (MLIP), accurately trained and tested using an on-the-fly Bayesian active learning approach~\cite{Bastonero2026}.

\begin{figure}
    \centering
    \includegraphics[width=.48\textwidth]{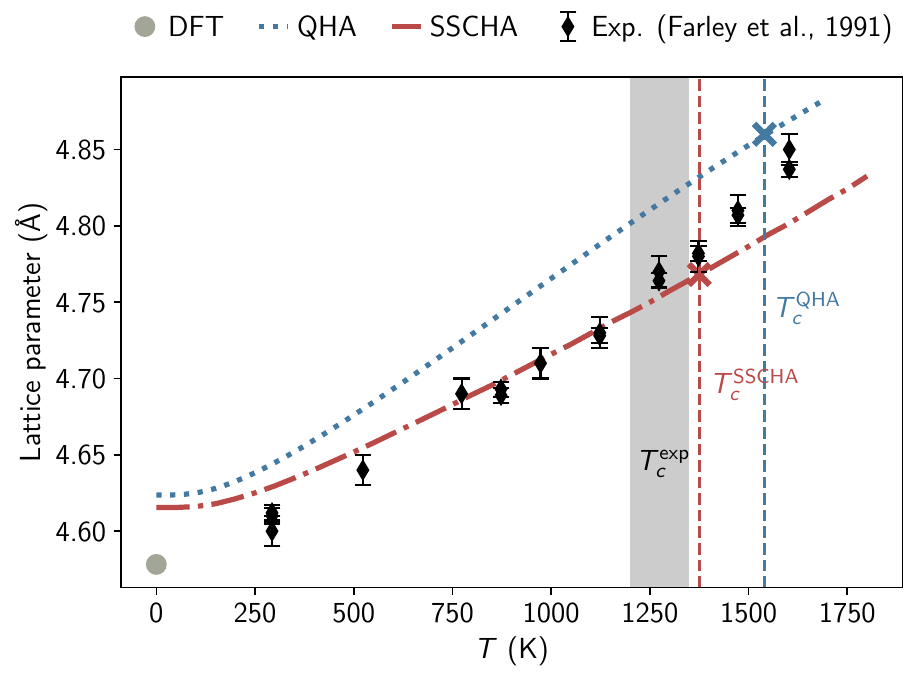}
    \caption{
        \textbf{\ce{Li2O} conventional cell lattice parameter as a function of temperature.}
        Comparison of lattice thermal expansion computed using QHA (blue dotted line) and SSCHA (red dashed line). The black diamonds represent the experimental values with their uncertainties from Ref. \citenum{Farley1991}. We highlight the temperatures where the phase transition occurs with vertical dashed lines and two crosses, for both SSCHA and QHA. The vertical shaded area represents the experimental T$_c$ range of the superionic transition.
    }
    \label{fig:thermalexpansion}
\end{figure}

\begin{figure*}
    \centering
    \includegraphics[width=.99\textwidth]{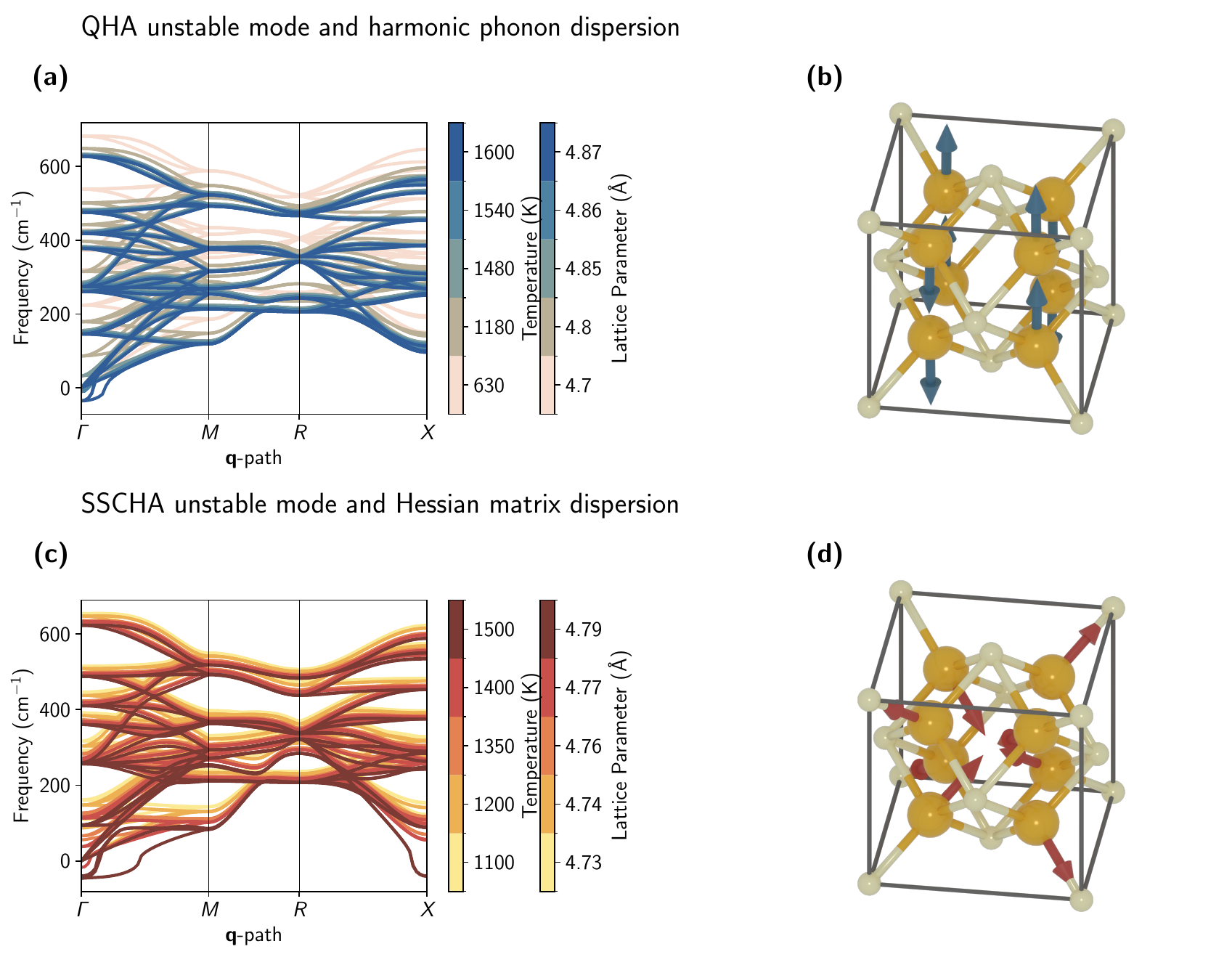}
    \caption{
        \textbf{\ce{Li2O} phonons as a function of temperature and lattice parameter:}
        (a) Harmonic phonon dispersion at different lattice parameters, with associated temperature calculated using the QHA. The instability arises in $\Gamma$ at $\approx 1540$~K, due to out-of-phase Li displacements represented in (b). (c) SSCHA phonon dispersion at different temperatures, with corresponding lattice parameters. The $\Gamma$ instability occurs at $\approx 1400$~K, with phonon eigenmodes reproduced schematically in (d).
    }
    \label{fig:phasetransition}
\end{figure*}

We start our analysis by computing the thermal expansion of \ce{Li2O}.  To understand the contribution of anharmonicity, we compare the same quantity calculated using SSCHA and QHA. For SSCHA, the equilibrium volume is obtained by relaxing both the volume and the atomic positions using the SCHA stress tensor~\cite{Monacelli2018}.
In Fig. \ref{fig:thermalexpansion}, we report the results and compare them to the experimental measurements in Ref.~\cite{Farley1991}. We note that QHA overestimates the thermal expansion, as already observed in previous works~\cite{Gupta2012, Nelson2018}. On the contrary, SSCHA accurately reproduces the thermal expansion in the majority of the experimental temperature range, between about 300 K and 1400 K.
Above 1400~K, the system undergoes to a structural phase transition (see below). SSCHA calculations retain the symmetry of the low-temperature \ce{Li2O} phase, which is no longer stable, and could be one cause of the deviation between the SSCHA volume from the experiment.
We calculate the transition temperatures with both methods, and we mark them in the plot. They are determined as the temperature at which a phonon becomes unstable (see Fig. \ref{fig:phasetransition} and later discussion). 
To determine the critical temperature T$_c$ using SSCHA, we calculate the Hessian of the positional free energy $F(\mathbf{R})$ (see Methods) at each $T$ and detect the onset temperature of the phonon softening.
We notice from Fig. \ref{fig:thermalexpansion} that, for the same volume, \emph{i.e.}, the same unit cell with a lattice parameter of 4.76 \AA, the harmonic phonons remain stable, whereas the SSCHA phonons already show an instability. This indicates that volume is not the only relevant factor driving this transition, and that anharmonic effects play a decisive role.

% --- THERMAL EXPANSION
% ---
%
In Fig.~\ref{fig:phasetransition}(a) and (c), we report the temperature-dependent phonon dispersion computed with QHA and SSCHA, respectively.
We observe the instability at the $\Gamma$ point, signalling the onset temperature of the structural phase transition. 
Within QHA, the instability occurs between 1480~K and 1540~K, while it occurs between 1350~K and 1400~K from the SSCHA calculations.
The experimetal superionic transition in Li$_2$O is commonly reported at approximately 1200~K, although estimates range from 1200 to 1350~K depending on the observable and study~\cite{Farley1991,Oishi1979,Hull1988,Strange1990}.
In Fig.~\ref{fig:thermalexpansion}, the experimental transition temperature is therefore reported with a shaded area representative of this range.
As for the volume, these results provide further evidence that anharmonicity is fundamental for accurately describing Li$_2$O around the superionic regime, which cannot be properly captured by harmonic methods, such as the quasi-harmonic approximation.
% 
% --- HARMONIC
% ---
%
Notably, we find that anharmonicity is so strong that it even renormalizes the eigenmode driving the instability.
In Fig. \ref{fig:phasetransition}(b) and (d), we depict the displacements associated to the unstable mode: 
The phonon instability arises from an optical mode at $\Gamma$ of the conventional 12-atom unit cell, and involves the out-of-phase displacements of Li ions only.
According to the QHA, the Li ions coherently move towards the same direction~\footnote{The displacement pattern in Fig. \ref{fig:phasetransition} (b) is plotted at \textbf{q}$\rightarrow$ 0 to better distinguish it from the degenerate second unstable mode in $\Gamma$. The latter represents the same displacement, but along the (110) plane.}. The same pattern was found in Refs.~\citenum{Fracchia1998, Goel2004, Gupta2012} using the harmonic and quasi-harmonic approximations.
Interestingly, when anharmonicity is accounted for, the picture changes.
The eigenvector now describes Li atoms moving toward an octahedral interstitial site (Fig.~\ref{fig:phasetransition}(d)), rather than toward a tetrahedral site as predicted by the QHA (Fig.~\ref{fig:phasetransition}(b)). Similarly, Wang and coauthors~\cite{Wang2015} identified a migration pathway in \ce{Li2S} in which Li atoms move between tetrahedral sites through an intermediate octahedral site.
Moreover, other works based on molecular dynamics on Li$_2$O and other superionic materials \cite{Jun2024,Wu2025,Goel2004,Gupta2019} also suggest that Li atoms jump from the tetrahedral positions to other tetrahedral positions, passing through the octahedral interstitial positions in the superconducting phase.
The neutron diffraction experiment by Farley et al.\citenum{Farley1991} likewise reported Li displacements toward the same $\langle 111 \rangle$ directions. 
These observations show that the phonon softening predicted by SSCHA is associated with the onset of the superionic phase transition, which provides a strong evidence of its microscopic mechanism. %gives the possible microscopical mechanism for it.

\begin{figure*}
    \centering
    \includegraphics[width=.9\textwidth]{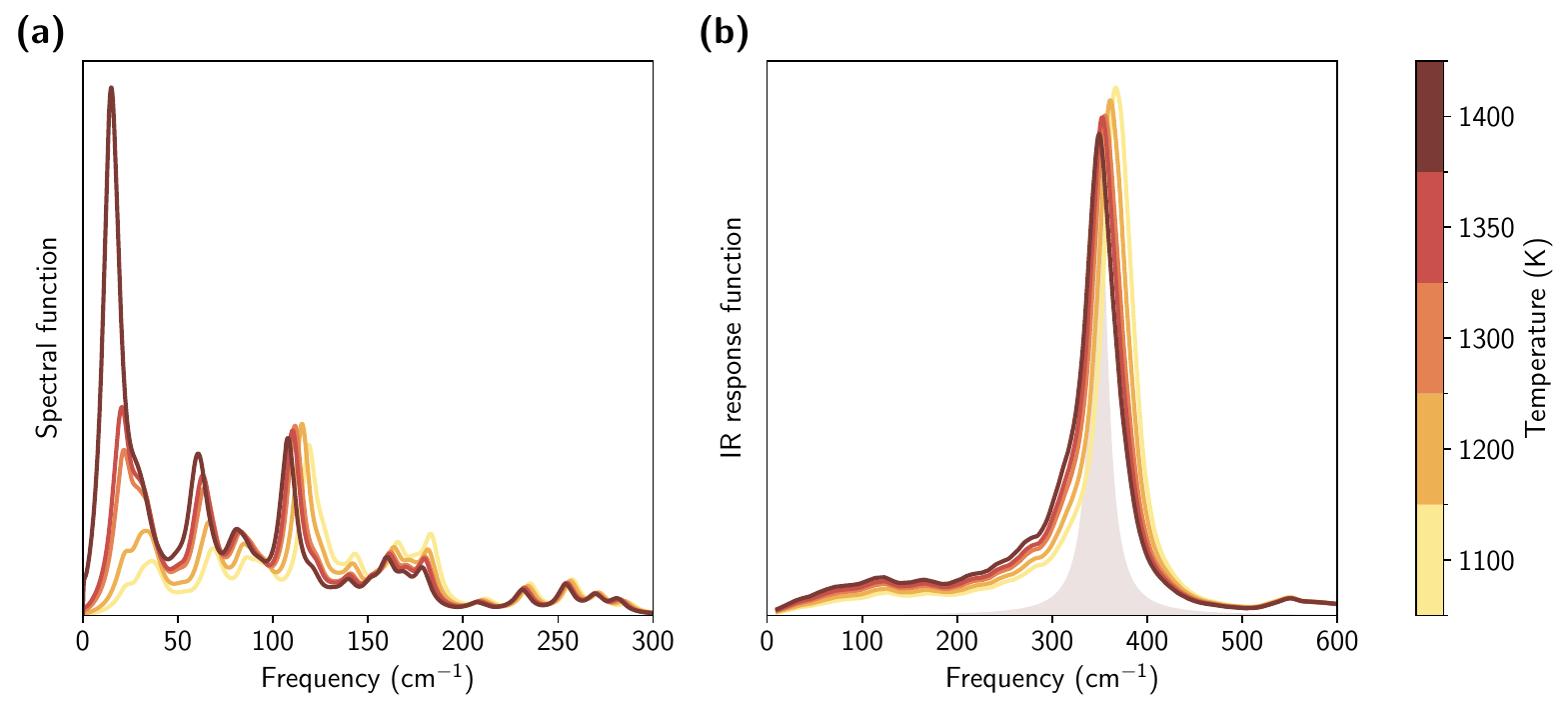}
    \caption{
        \textbf{\ce{Li2O} Spectral function and IR spectrum:}
        (a) Spectral function of the lowest optical mode (A mode) at \textbf{q} = $\Gamma$, zoom at low frequencies. (b) IR response function. The shadowed area represents a Lorentians function centered at the dominant pole of the Green's function at 1400 K, with a smearing of 10 Ry. %\textbf{\color{blue} forse potremmo metterli tutti i poli, rinormalizzati per il loro peso?}}
        }
    \label{fig:spectral_funct}
\end{figure*}

Finally, we analyze the spectral function $\sigma$($\omega$) and the IR response of \ce{Li2O} at different temperatures near the phase transition (1100 K--1400 K). To do that, we use the time-dependent self-consistent harmonic approximation (TD-SCHA)~\cite{Monacelli2021, Siciliano2023}. TD-SCHA is a dynamical theory that provides direct access to the
spectral functions, capturing the effects of SCHA phonon--phonon interactions. If the quasiparticle picture holds, SCHA phonons are represented by spectral functions of Lorentzian-shape with finite widths and shifted frequencies.
When anharmonicity is sufficiently strong, the quasiparticle picture can break down, and the spectral function exhibits more complex, non-Lorentzian features. This is the case of Li$_2$O, as we see from the results in Fig. \ref{fig:spectral_funct}.
In Fig. \ref{fig:spectral_funct}(a), we show the spectral function of the lowest-frequency optical mode at $\Gamma$. According to our previous Hessian calculation, this mode lies at 109 cm$^{-1}$ at T = 1100 K; the mode softens when increasing temperature, finally becoming unstable at the critical temperature (see Fig. \ref{fig:phasetransition}(c)).
The spectral function $\sigma$($\omega$) of this mode, on the other hand, reveals multiple peaks - \emph{satellites} - in the frequency range between 0 and 200 cm$^{-1}$.
Going towards the phase transition, the peak at around 0 cm$^{-1}$ increases significantly.
Notably, the spectral function is proportional to the signal probed in inlastic neutron or X-ray scattering, providing a way to identify the phase transition experimentally.
The IR spectrum is shown in Fig. \ref{fig:spectral_funct}(b). A main peak around 350 cm$^{-1}$ is well visible, which corresponds to the IR mode obtained from the SSCHA Hessian calculations at 316 cm$^{-1}$ for T = 1100 K and 271 cm$^{-1}$ for T = 1400 K. When SCHA phonon--phonon interactions are included, the main IR peak shifts to 365 cm$^{-1}$ at 1100 K and 349 cm$^{-1}$ at 1400 K.
Moreover, the IR response deviates from the quasi-particle picture, in which the spectrum is described by a Lorentzian centered at the shifted frequencies.
For comparison, we plot in gray the Lorentzian centered at pole of the Green function at 1400 K with the the highest-intensity. 
The calculated IR peak is asymmetric, and the low-energy region is dominated by satellite features between 0 and 200 cm$^{-1}$.
This distinctive asymmetric shape may provide an experimental fingerprint of the phase transition through infrared spectroscopy.
Interestingly, these low-energy features may suggest a possible route to experimentally inducing the phase transition, motivating future experiments in which appropriately tuned THz pulses could selectively drive the associated  dynamics~\cite{Libbi2025}.

%============Conclusion==================
In conclusion, we employed SSCHA to study the prototypical fast-ion conductor Li$_2$O. We find that properly including anharmonicity  reproduces the material's thermal expansion more accurately than the quasi-harmonic approximation, and that the temperature at which the phonon softening occurs using SSCHA agrees well to the experimental superionic transition temperature.
These findings establish the large impact of anharmonicity on the Li$_2$O superionic phase transition.
We demonstrate that the main mechanism behind the transition cannot be explained by volume expansion alone, unlike previously thought, but is strongly associated with anharmonicity.
Notably, the transition is signaled by a softening of the Li-ion displacements toward octahedral interstitial sites rather than tetrahedral sites, as predicted from harmonic theory. Octahedral sites  were previously proposed as intermediate positions along the lithium diffusion pathway during the fast-diffusion regime. 
The agreement of both the critical temperatures and the soft-mode eigenvectors with previous experiments and MD simulations indicates that the softening of the free-energy Hessian corresponds to the superionic phase transition. We capture this transition for the first time using a lattice-dynamical approach, revealing its microscopic mechanism.
Finally, we investigate the spectral properties through the time-dipendent SCHA, revealing that \ce{Li2O} is not only highly anharmonic, but that its anharmonicity is strong enough to break the quasiparticle picture.
This conclusion may also apply to other superionic conductors, which would be interesting to investigate further.
This study is a step toward understanding the general mechanisms behind the superionic regimes in terms of lattice dynamics.
We show how SSCHA is a useful framework to investigate superionic materials, paving the way for the efficient and reliable design of next-generation solid-state electrolytes.  

%TC:ignore
\section{Methods}
The calculations presented in this work were performed using the \texttt{python-sscha} code~\cite{Monacelli2021a}, with the MLIP trained using the on-the-fly Bayesian approach presented in our previous work \cite{Bastonero2026}. 
The ML training only required 44 DFT calculations, carried out using the PBEsol exchange-correlation functional and the \textsc{Quantum ESPRESSO} package \cite{Giannozzi2009}. The pseudopotentials are taken from the SSSP Library PBEsol precision library (version 1.3) \cite{Prandini2018}.
The SSCHA phonon dispersions are calculated taking the eigenvalues (and corresponding eigenvectors for the displacement pattern) of the Hessian of the positional free energy, \emph{i.e.}, the second-order derivative of the free energy with respect to the atomic coordinates \textbf{R}:
\begin{equation}
H_{ab} = \frac{1}{\sqrt{ M_a M_b}} \cdot \frac{\partial^2 F}{\partial R_a \, \partial R_b}.
\end{equation} 
This can be seen as the natural generalization of the harmonic dynamical matrix, but now including thermal and quantum nuclear effects. 
We exploit the analytic formulation from Ref. \citenum{Bianco2017}, and we include up the third-order term. The thermal expansion was calculated with a 2x2x2 supercell with 5.000 configurations in the ensemble, while for the Hessian calculations we used an ensemble of 100.000 configurations in the same supercell.

The same MLIP was employed for the QHA calculations, where the harmonic force constants are computed in $2\times2\times2$ supercells (96 atoms) using the finite-displacement method, as implemented in the \texttt{Phonopy} package~\cite{Togo2015,Togo2023,Togo2023a} with a default displacement of 0.01~$\AA$. The phonons are then interpolated into finer meshes of $21\times21\times21$ to compute the harmonic free energies required for the QHA calculations.

The spectral function and the IR absorption were computed using the time-dependent SCHA in its linear response formulation\cite{Monacelli2021,Siciliano2023}. We used the \texttt{tdscha} Python package within the modified branch with SCHA commit \texttt{dc4bbdd1d73\-26ab02810b\-996c8c57782\-0d82f8d2}. We used an 8x8x8 interpolation grid and the Lanczos algorithm, and we include the three-phonon scattering (bubble approximation).
The spectral function is computed from the diagonal elements of the dynamical one-phonon Green function: 
\begin{equation}
\sigma(\mathbf{q}, \omega) = - \frac{\omega}{\pi}\Tr \text{Im} \mathbf G(\mathbf{q},\omega).
\end{equation}
The phonon Green function is given by:
\begin{equation}
    \mathbf G= \mathbf G^{(0)} + \mathbf G^{(0)}\mathbf \Sigma \mathbf G,
\end{equation}
with \textbf{G}$^{(0)}$ is the Green function of the non-interacting SCHA phonons, and $\mathbf\Sigma$ their self-energy taking into account the interactions. 
We refer to Refs. \citenum{Monacelli2021,Siciliano2023} for details on the method.

The IR response function is obtained as the imaginary part of the polarizability, linked to the interacting one-phonon Green function through the formula:
\begin{equation}
    \chi(\omega) =\sum \frac{Z_{a\alpha}Z_{b\beta}}{\sqrt{m_am_b}}G_{ab}(\omega) ,
\end{equation}
where $Z_{a\alpha}$ are the Born effective charges for the $a$-atom coordinates with polarization $\alpha$.
The Green functions are calculated using a frequency broadening for the continued fraction of $\eta$ = 10 Ry.
The polarization is (1,0,0).

\section*{Code availability}
All the code employed in this work are open source and available on  GitHub: \texttt{python-sscha} distribution (\url{https://github.com/SSCHAcode/python-sscha}), the \texttt{tdscha} (\url{https://github.com/SSCHAcode/tdscha}) and FLARE code~\cite{Vandermause2020, Vandermause2022} (\url{https://github.com/mir-group/flare}).
All the data will be made available in the Materials Cloud Archive upon publication.

\begin{acknowledgement}

L.B., C.C. and N.M. gratefully acknowledge support from the Deutsche Forschungsgemeinschaft (DFG) under Germany’s Excellence Strategy (EXC 2077, No. 390741603, University Allowance, University of Bremen) and Lucio Colombi Ciacchi, the host of the “U Bremen Excellence Chair Program”.
L.B., C.C., and N.M. also acknowledge funding by MaX “Materials Design at the Exascale”, a Center of Excellence co-funded by the European High Performance Computing Joint Undertaking (JU) and participating countries under grant agreement No. 101093374.
We acknowledge support by the NCCR MARVEL, a National Centre of Competence in Research, funded by the Swiss National Science Foundation (Grant number 205602). 

\noindent\textit{Conflict of Interest}: The authors declare no competing financial interest.

\end{acknowledgement}
%TC:endignore
%------------------------------------------------------------------------------------ %
%------------------------------------------------------------------------------------ %

\bibliography{biblio}
\end{document}